\documentclass{ws-procs961x669}

\newcommand{\ket}[1]{|{#1}\rangle}

\begin{document}

\title{Opportunities and Challenges in Fault-Tolerant Quantum Computation}

\author{Daniel Gottesman}

\address{Joint Center for Quantum Information and Computer Science (QuICS) and Computer Science Department, University of Maryland, College Park, MD 20742, USA \\
E-mail: dgottesm@umd.edu}
\address{Keysight Technologies, Waterloo, ON, Canada}

\begin{abstract}
I will give an overview of what I see as some of the most important future directions in the theory of fault-tolerant quantum computation.  In particular, I will give a brief summary of the major problems that need to be solved in fault tolerance based on low-density parity check codes and in hardware-specific fault tolerance.  I will then conclude with a discussion of a possible new paradigm for designing fault-tolerant protocols based on a space-time picture of quantum circuits.
\end{abstract}


\bodymatter

\section{Introduction}
\label{sec:intro}

Building a large quantum computer is a daunting task.  One of the main obstacles to doing so is errors.  There are two reasons why errors are expected to be a more serious problem for quantum computers than for classical computers: First, quantum computers are necessarily made from extremely small components that can behave quantumly, and small components are always going to be more sensitive to disturbances than larger components.  Second, quantum states are susceptible to more types of errors than classical computers; in particular, decoherence can be caused by any leakage of information about the state of the system into the environment, which could be through a full-fledged measurement or just a single atom wandering by and becoming correlated with the quantum system.  Currently, there is a lot of interest in algorithms for the NISQ (Noisy Intermediate-Scale Quantum) era,\cite{NISQ} in which modest-sized quantum computers without error correction are used to solve a restricted set of problems, but we expect that ultimately, to build larger universal quantum computers, they will need to be fault tolerant.

So what is fault tolerance?  Fault tolerance is a method of transforming quantum circuits into new circuits that involve extra qubits and more gates but are robust against a low level of errors occurring throughout the computation.  This means that all the quantum components --- state preparation, gates, and measurements --- are susceptible to errors.  

Fault tolerance should be distinguished from quantum error correction, which is by itself only suitable for communication and memory scenarios.  In quantum error correction, Alice encodes some qubits that need to be protected by adding extra qubits and performing a unitary encoding operation.  The qubits of the code are sent through a quantum channel (such as a communications channel or natural decoherence processes when the qubits are stored for some time) to Bob, who then decodes and uses the properties of the quantum error-correcting code (QECC) to identify and correct any errors that occurred in the quantum channel.  Quantum error correction assumes that Alice's encoding and Bob's decoding procedures are perfect.  For a fault-tolerant protocol, we need to remove this assumption and find ways of creating encoded states, correcting errors, and performing gates on the encoded qubits that continue to work even though the quantum gates used to do these procedures are themselves imperfect.  A fault-tolerant protocol does encode the qubits in a QECC, but adds on top of that an encoding of all the circuit elements that make up a quantum computation.

I shall begin by giving an overview of the current state of the art in fault tolerance in Sec.~\ref{sec:current}, along with some useful definitions, then move on to discuss some specific current and future research directions on low-density parity check codes (Sec.~\ref{sec:LDPC}) and hardware-specific fault tolerance (Sec.~\ref{sec:hardware}), and then conclude with a discussion of a possible new paradigm for designing fault-tolerant protocols in Sec.~\ref{sec:spacetime} and Sec.~\ref{sec:stspecifics}.  

This paper is not intended to be an exhaustive list of open problems relating to fault tolerance, or even a prioritization of which open problems I feel deserve the most attention, but is instead a focus on three specific (albeit broad) directions that I think will be important in coming years.  Certainly there will also be a continuing need to optimize existing protocols, with magic state distillation\cite{magicstates} a likely focus.  And I expect further experimental progress in demonstrating ever-larger and more capable fault-tolerant systems.  Quantum error correction has proven to be quite useful in a variety of other areas beyond simply building quantum computers, and likely more such applications will be discovered, with fault-tolerant techniques starting to play a larger role as well.

\section{Current State of the Art}
\label{sec:current}

A fault-tolerant quantum protocol\cite{ShorFT,chapter} is a mapping from quantum circuits that we wish to perform to larger circuits that are fault tolerant.  The qubits of the original circuit are referred to as \emph{logical qubits} and the gates of the original circuit are \emph{logical gates}, whereas the qubits and gates of the fault-tolerant circuit are \emph{physical qubits} and \emph{physical gates}.

In the current paradigm of design for fault-tolerant protocols, a fault-tolerant mapping replaces logical qubits with qubits encoded in a QECC and each circuit element (state preparation, gate, and measurement) is replaced by an appropriate \emph{fault-tolerant gadget}.  I will use the term ``gadget'' in a slightly broader way to describe an indivisible unit of a fault-tolerant protocol, not necessarily performing any single logical circuit element.  I will not give a precise definition of fault tolerance in the standard paradigm, but under current design principles, the main goal is for the gadget to keep \emph{error propagation} under control.  Error propagation is what happens when a two-qubit gate acting correctly interacts two qubits, one of which has experienced an error.  After the gate, it is often the case that the state now has a two-qubit error relative to the ideal world with no errors.  For instance, if there is a bit flip ($X$) error on the first (control) qubit and then a perfect CNOT gate, now both qubits have bit flip errors, as the second qubit gets flipped exactly when it is not supposed to be flipped.  If both qubits are physical qubits in the same QECC, this is likely to be a problem.  For instance, if the code is able to correct arbitrary single-qubit errors, it was able to correct the error on the state before the gate but not after the gate.

The easiest way to avoid error propagation is to use \emph{transversal} gates.\cite{ShorFT}  The simplest example of a transversal gate is a tensor product of single-qubit gates, which certainly cannot propagate errors.  Transversal multiple-qubit gates interact more than one block of the QECC, each of which is a separate copy of the QECC in use encoding different logical qubits.  For such a gate to be transversal, it must interact only corresponding qubits from the different blocks, i.e., the 1st qubit of the first block interacts with the 1st qubit of the second block, the 2nd qubit of the first block interacts with the 2nd qubit of the second block, and so on.  However, it is not possible to perform a universal set of gates using just transversal gates,\cite{EastinKnill} so additional more complicated techniques are needed, which I will not discuss in this paper.

The choice of an appropriate QECC is central to the design of a fault-tolerant protocol.  One large and widely-used family of QECCs are the \emph{stabilizer codes}.~\cite{chapter,stabilizer,CRSS} 
Stabilizer codes are defined by a set of constraints (generators of a \emph{stabilizer group} $S$), which require valid codewords to be a $+1$ eigenstate of elements of the \emph{Pauli group}, which consists of tensor products of Pauli operators $I$, $X$, $Y$, $Z$,
\begin{equation}
X = \begin{pmatrix} 0 & 1 \\ 1 & 0 \end{pmatrix}, \ Y = \begin{pmatrix} 0 & -i \\ i & 0 \end{pmatrix}, \ Z = \begin{pmatrix} 1 & 0 \\ 0 & -1 \end{pmatrix}.
\end{equation}
with an overall phase of $\pm 1$, $\pm i$.  The stabilizer group elements must commute so that they have simultaneous $+1$ eigenstates.

Stabilizer codes are often characterized in terms of three parameters in the notation $[[n,k,d]]$.  $n$ is the number of physical qubits and $k$ is the number of logical qubits, which for a stabilizer code is equal to $n - r$, with $r$ being the number of independent generators of the stabilizer $S$.
$d$ is the \emph{distance}, which for a stabilizer code is the smallest weight (number of non-trivial Pauli factors) of a Pauli group element $F$ for which $F$ commutes with all elements of $S$ but is not itself in $S$.  Let $N(S) = \{ P | P \text{ is a Pauli and } PM = MP \ \forall \, M \in S \}$.  Then the distance is the lowest weight of an element of $N(S) \setminus S$.  The distance is the minimum number of qubits that must be touched in order to change one logical codeword to a different logical codeword.  $d$ addresses the ability of the QECC to correct errors, and a code with distance $d$ can correct arbitrary errors affecting up to $\lfloor (d-1)/2 \rfloor$ physical qubits.  This is because an error $E$ that anticommutes with an element $M$ of the stabilizer changes the eigenvalue of the codeword from $+1$ for $M$ to $-1$.  Thus, measuring the eigenvalues of the generators of the stabilizer gives us a binary vector of length $n-k$ called the \emph{error syndrome}, which can be used to identify which error occurred.

In Sec.~\ref{sec:spacetime}, I shall need a generalization of stabilizer codes called \emph{subsystem} stabilizer codes.\cite{Poulingauge}  One way to think about subsystem codes is as a QECC where we do not care about the value of some of the logical qubits and that errors that only change those logical qubits are not considered a problem.  A stabilizer subsystem code has a stabilizer group like a regular stabilizer code but also a \emph{gauge} group $G$ also consisting of elements of the Pauli group.  $G$ does not need to be Abelian, but $S \subseteq G \subseteq N(S)$.  $G$ represents changes to the ``unimportant'' logical qubits.  The distance of the subsystem stabilizer code is then the weight of the smallest $F$ that commutes with all elements of $S$ but is not in $G$, i.e., the minimum weight of an element of $N(S) \setminus G$.

I also want to add one more element, which has not previously been discussed in the literature as far as I know.  We will consider some elements of the stabilizer to be \emph{masked}, meaning that while correct codewords are constrained to be $+1$ eigenstates of the masked elements of the stabilizer, we are unable to measure the eigenvalues of those elements for whatever reason.  Masking is helpful when considering QECCs with geometric constraints on measurements,\cite{LDPClocal} will be needed in Sec.~\ref{sec:spacetime}, and may have other applications as well.  Formally, let us define two additional subgroups $U$ and $T$, $U \subseteq T \subseteq S$.  $U$ will be the \emph{always unmasked} subgroup of stabilizer elements whose eigenvalues can always be measured.  $T$ will be the \emph{temporarily masked} subgroup of stabilizer elements whose eigenvalues can be possibly be measured at some point in the future.  $T \setminus U$ are stabilizer elements which cannot be measured currently but might be available later.  $S \setminus T$ contains those stabilizer elements whose eigenvalues can never be measured, \emph{permanently masked} elements.  They are still relevant because an error that produces on net an element of $S \setminus T$ will leave codewords unchanged.  The permanently masked stabilizer elements differ from gauge elements in two respects: First, acting by a masked stabilizer element leaves a codeword unchanged, whereas a gauge element can change the state, albeit in an unimportant way.  Second, a gauge element is paired with another gauge element that anticommutes with it, whereas masked stabilizer elements commute with everything in the gauge group.  We can define distances $d_U$, the minimum weight of an element of $N(U) \setminus S$ (or $N(U) \setminus G$ if it is a subsystem code), and $d_T$, the minimum weight of an element of $N(T) \setminus S$ (or $N(T) \setminus G$).  Note that $d_U \leq d_T \leq d$.  $d_U$ and $d_T$ encode the QECC's ability to correct errors using only information from the unmasked generators or with the temporarily masked generators but without the permanently masked generators.

Stabilizer codes have a special relationship with a group of unitary gates known as the \emph{Clifford group}.\cite{GFT, Heisenbergrep} The Clifford group is defined as the set of unitaries $U$ such that when any element $P$ of the Pauli group is conjugated by $U$, the result $UPU^\dagger$ is another element of the Pauli group.  Encoding and decoding circuits for stabilizer codes can be done using just the Clifford group, and yet circuits consisting of just Clifford group elements can be efficiently simulated on a classical computer.  This is one reason why stabilizer codes are useful.  Another reason is that for Clifford group elements, it is easy to understand the behavior of error propagation: Pauli errors $P$ before the Clifford group gate $U$ propagate to $UPU^\dagger$ after the gate.  For instance, the Hadamard rotation $H$ is an element of the Clifford group and it propagates $X$ to $Z = HXH$ and $Z$ to $X = HZH$.  The CNOT gate, another element of the Clifford group, propagates
\begin{align}
X \otimes I  &\rightarrow X \otimes X \\
I \otimes X  &\rightarrow I \otimes X \\
Z \otimes I &\rightarrow Z \otimes I \\
I \otimes Z  &\rightarrow Z \otimes Z.
\end{align}
The propagation of other elements of the Clifford group can be understood just by multiplying the relations on $X$ and $Z$ together appropriately.  For instance, $Y = iXZ$, so CNOT maps $X \otimes Y \rightarrow i (X \otimes X) (I \otimes X) (Z \otimes Z) = Y \otimes Z$.

In order to protect a large quantum computation, it is not enough to focus on fault tolerance with just a single small quantum code.  For fixed $d$, there will always be a chance of a cluster of errors randomly occurring that overwhelms the capacity of the code to correct those errors.  Therefore, if we wish to protect very large computations, we need not just a single code, but a family of larger and larger codes with a distance that grows with the size of the code.  If there are $n$ physical qubits and a probability of $p$ of error per qubit per time step, then we expect there to be about $pn$ errors at every time step.  This might suggest that we need a distance of at least $2pn$, but in fact, we can get away with a significantly smaller (including sub-linear) distance.  This is because the distance captures the code's potential to correct the \emph{worst-case} error, but randomly-located errors are unlikely to conspire in the worst possible ways.  This is convenient, and there are a number of families of QECCs that allow us to correct \emph{typical} errors in the limit as $n \rightarrow \infty$ even though they have distances that scale slower than $n$.

When we put one of these families of QECCs together with an appropriate protocol for fault tolerance, we get the \emph{threshold theorem}:\cite{KLZ,AB,KitFT,AGP}
\begin{theorem}
There exists a threshold value $p_t$ with the following property: If the error rate $p$ per physical gate or time step is below $p_t$, then for any $\epsilon > 0$, there exists a fault-tolerant protocol such that any logical circuit of size $T$ is mapped to a circuit with $\mathrm{polylog(T/\epsilon)}$ times as many qubits, gates, and time steps, and the output of the fault-tolerant circuit is correct except with probability $\epsilon$. 
\end{theorem}
Here, $\mathrm{polylog(T/\epsilon)}$ means a polynomial in the logarithm of $T/\epsilon$.
This statement of the threshold theorem skips some underlying assumptions, such as the nature of the errors, but in fact many of those assumptions can be relaxed, and there is still provably a threshold for a very wide variety of weak local error models, including many non-Markovian error sources.\cite{AGP} The threshold theorem is critical for the experimental realization of large-scale quantum computers, since it says that experimentalists have a constant target value $p_t$ for their error rates, and won't need to continue to improve their gate fidelities in order to make larger and larger quantum computers.  Instead, once the qubits are accurate enough, building a big quantum computer is solely a question of adding more physical qubits with the same level of reliability.

However, the statement of the threshold theorem is also somewhat misleading in practice, since it suggests that there is a single magical number $p_t$ that is the target for all efforts to build a quantum computer, which is not the case.  In fact, the precise numerical value of the threshold is sensitive to assumptions about the system, including both the details of the error model and the specific fault-tolerant protocol in use.  Simulations suggest the threshold can be made as high as a few percent in a depolarizing error model,\cite{Knillhighthresh} in which each qubit has an error at each time step with probability $4/3 p$, and if there is an error, the qubit is completely randomized.  (I have included the factor of $4/3$ because when the qubit is randomized, there is a $1/4$ chance that the resulting error is the identity and the qubit doesn't change.) However, to achive such a high threshold value, we need ridiculously high overheads (e.g., $10^9$ physical qubits per logical qubit), so these protocols are not practical.

Another simplification implicit in the statement of the threshold theorem is that there is just a single relevant parameter needed to quantify the amount of error, but in real systems, each different physical gate will have a different error profile associated with it, so the error model actually involves many parameters.  The logical gates will have different effective error rates and also different kinds of errors than the corresponding physical gates.  In this context, the threshold theorem is still valid, but instead of the threshold being a single number, it is instead a hypersurface in a high-dimensional space.  Points inside the hypersurface have their error operators driven towards the identity, whereas points outside the hypersurface are mapped to worse error models.  This is because the extra overhead involved in a fault-tolerant protocol results in additional opportunities for errors to occur.  A fault-tolerant protocol is a race between correcting errors and the new errors being constantly created as the circuit proceeds, and the threshold surface demarcates the balance point between these two influences.  Outside the threshold, the protocol cannot correct errors as fast as they occur, and so more gates will cause the amount of errors corrected to fall even further behind the number of errors occurring.

One fault-tolerant protocol that has emerged as being of considerable practical relevance is fault tolerance based on the family of \emph{surface codes}.\cite{DKLP,RH,Fowler} Fault tolerant protocols using surface codes have a high threshold error rate, about $0.7\%$ for depolarizing noise, and can be easily arranged in a two-dimensional architecture with nearest-neighbor physical gates.  Their overhead is still a bit high at hundreds or thousands of physical qubits per logical qubit, but if necessary, we can tolerate this much overhead if it is the only way to build a quantum computer.

\begin{figure}
\begin{center}
\begin{picture}(140,140)

\multiput(20,20)(20,0){6}{\line(0,1){100}}
\multiput(20,20)(0,20){6}{\line(1,0){100}}

\put(10,84){\makebox(12,12){$Z$}}
\put(30,84){\makebox(12,12){$Z$}}
\put(24,68){\makebox(12,12){$Z$}}
\put(24,100){\makebox(12,12){$Z$}}

\put(64,60){\makebox(12,12){$X$}}
\put(84,48){\makebox(12,12){$X$}}
\put(79,64){\makebox(12,12){$X$}}
\put(69,44){\makebox(12,12){$X$}}

\end{picture}
\end{center}
\caption{The surface code}
\label{fig:surface}
\end{figure}
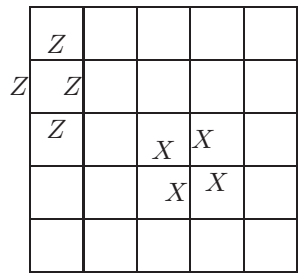

Surface codes are stabilizer codes and the constraints for a surface code are defined by a two-dimensional graph, frequently a square lattice, as in Fig.~\ref{fig:surface}.  The qubits are located on the edges of this graph.  For each face of this graph, the stabilizer has a generator which is a product of $X$ over each qubit on an edge bordering that face, and for each vertex, a generator which is a product of $Z$ over each qubit on an edge ending at that vertex.  The graph can be on a non-trivial two-dimensional manifold, such as a torus, but it is more practical to set appropriate boundary conditions at the edges of the surface, including possibly leaving holes in it, to create a code with the desired number of logical qubits.

\section{Low-Density Parity Check Codes}
\label{sec:LDPC}

As an alternative to surface codes, I have championed the idea of using high-rate low-density parity check (LDPC) codes for fault tolerance.\cite{LDPC} A quantum LDPC code is a stabilizer code whose generators have the following two properties:
\begin{itemlist}
\item Each generator acts on only a constant number of qubits
\item Each qubit is involved non-trivially in only a constant number of generators
\end{itemlist}
Here constant means constant for a family of codes when the number of physical qubits $n$ gets large.  Surface codes are an example of LDPC codes, but they have a low \emph{rate} $k/n$, the ratio of logical qubits to physical qubits.  However, there are other LDPC codes which have a high rate, even constant for large $n$, which are capable of correcting as many or more errors as a large surface code.  Such codes can in principle remove the polylogarithmic overhead from the threshold theorem, allowing a fault-tolerant protocol with constant qubit overhead.\cite{LDPC}

The family of all LDPC codes is very broad, so we should ask which specific subset of LDPC codes is the most interesting for the purpose of fault tolerance.  Currently, codes based on the hypergraph product construction~\cite{TZ09} seem promising, particularly expander codes.\cite{LTZ15} Expander codes feature a distance that grows as $\sqrt{n}$, an efficient decoding algorithm that works for typical errors even in the fault-tolerant context (when faults can occur while performing error correction),\cite{FGL18} and a threshold of about $7.5\%$ in the non-fault-tolerant context,\cite{GGKL20} which is slightly worse than the surface codes, but not dramatically so.

It is also worth noting that there has been spectacular progress over the last few years in the construction of quantum LDPC codes,\cite{goodLDPC,Tanner,GPT22,DHLV22} finally resolving the long-standing open problem of whether there exist families of \emph{good} LDPC codes, namely code families with parameters $[[n,k,d]]$ with $d/n$ and $k/n$ both constant as $n \rightarrow \infty$.  Such codes can correct many more errors in the worst case than a hypergraph product code, but it is not yet clear whether the new LDPC code constructions can improve our fault-tolerant protocols based on LDPC codes. 

LDPC codes have the potential to reduce overheads relative to surface codes by an order of magnitude or more,\cite{TDB21,CKBB21} but the work of making practical protocols with high-rate LDPC codes has only just begun.  In the fully fault-tolerant context, we know of protocols with a threshold of just under $0.3\%$,\cite{TDB21} which again is slightly worse than surface codes but perhaps good enough given the potential gain in overhead.  It also worth noting that we are still just at the beginning of understanding LDPC codes, so further improvement may be possible to match or even exceed the threshold of surface-code-based protocols.  Another advantage of high-rate LDPC codes over surface codes is that they have what is known as ``single-shot'' decoding, which means that the error correction procedure can be completed much more quickly, indeed in a time that is constant even as the code gets larger, whereas surface codes require a longer time for larger codes.

Still, there is a lot to be done if we wish to replace surface codes with high-rate LDPC codes.  One particular area where progress is needed is in how to construct fault-tolerant gates between logical qubits encoded in an LDPC code.  There are some general techniques that can be applied,\cite{LDPC} but these are rather inefficient and not very practically useful.  There are a number of recent papers investigating gate constructions for LDPC codes,\cite{CKBB21,KP19,BB22,QWV22} but none of the existing solutions is completely satisfactory.

One particular flaw in existing gate constructions is that they require that we perform the logical circuit in a sequential fashion, which means one gate at a time.  This is not an extra cost if the original circuit was already very sequential, but if the logical circuit to be performed is amenable to being parallelized, it would be a shame if a fault-tolerant version of it couldn't retain that benefit.  A recent result\cite{timeoverhead} using a non-LDPC family of codes shows that it is possible in principle to have a threshold for fault tolerance with constant overhead in the limit of large $n$ and a sub-polynomial time slowdown relative to even a highly-parallelized logical circuit.  It may be possible to find other gate gadget constructions which retain the other benefits of LDPC codes and additionally have a similarly low time overhead.

However, there is one inherent drawback to high-rate LDPC codes which is hard to circumvent.  Because these codes require a high connectivity in order to rapidly spread out the information in their many logical qubits, LDPC codes with a non-vanshing rate cannot be laid out so that all stabilizer generators are geometrically localized in two dimensions, or indeed any finite dimension.\cite{BK21a,BK21b,DBT21} This means that high-rate LDPC codes are most suitable for hardware platforms which allow long-range gates with little or no extra cost.  It may also be possible to lay out a fault-tolerant LDPC code-based protocol in such a way that only a handful of long-range gates are needed during the protocol, or even none at all, even though the stabilizer generators themselves are not all localized.\cite{LDPClocal}  We can draw inspiration from concatenated quantum codes here, which are also highly non-local but can still be arranged in a 2D or even 1D architecture with a fault-tolerant threshold.\cite{AB,Glocal}

\section{Hardware-Specific Fault Tolerance}
\label{sec:hardware}

Another important route for fault tolerance research is to take into account specific properties of the hardware platform being used.  To some extent this is already done, for instance in the use of geometrically local gates or not, but there is much more that can be done in this respect.

One avenue is to take advantage of the full Hilbert space of the platform.  This is done, for instance, in the study of bosonic codes.\cite{GKP,cat1,cat2,binomial,survey} Systems with a continuous-variable degree of freedom are not uncommon, typically harmonic oscillators (at least, approximately harmonic) of some sort, and they present both interesting new opportunities and new challenges.  It is not possible to encode a continuous-variable system in another with full fault tolerance given the range of possible errors,\cite{HK21} but what \emph{can} be done is to encode a qubit in the bosonic mode fault tolerantly.  Control of these systems is experimentally challenging, since the encoding invariably involves some sort of non-linear process.\cite{noGaussian} However, a number of such codes have been realized in recent years.\cite{OPH+16,HMC+19,CET+20,FNM+19}

One advantage of using a bosonic code is that it can provide some degree of error correction immediately, leading to qubits that are more reliable than would be achieved by a simpler but non-error-correcting encoding of a qubit in the mode.  Unlike a code encoding qubits in qubits, this doesn't result in any increase in the number of modes used since the extra degrees of freedom in the mode would normally remain unused.  A bosonic code encoding one qubit in one mode will only have a limited amount of error-correction capability and therefore will not, by itself, be sufficient for a large quantum computer.  The usual course is to concatenate: Use a family of qubit codes with a threshold, such as surface codes or LDPC codes, and each physical qubit of that code is further encoded as the logical qubit of a bosonic code.\cite{Terhalboson} Because the bosonic code has a large physical Hilbert space, it gives some information about whether its own error-correction procedure is likely to have succeeded (when the state is close to a correct codeword) or failed (when the state is far from a correct codeword), and this information can help decode the qubit code at higher error rates than it would normally tolerate.  One worthwhile approach that has not been explored enough is to find bosonic codes that use multiple modes without requiring concatenation.\cite{PHboson}

Another important goal should be to develop fault-tolerant protocols that take advantage of as much information about the errors as possible.  Standard protocols are designed and analyzed for simplistic error models, usually the depolarizing channel.  There are two reasons for this.  One is that it is hard to simulate general error models, whereas Pauli errors like the depolarizing channel can be efficiently simulated classically as they propagate through Clifford group circuits.\cite{Heisenbergrep} The second is that error propagation through circuits changes not just the location and number of errors, but their type.  As a simple example, suppose a qubit has a phase error ($Z$) and then undergoes a Hadamard gate.  The gate is perfect but now the qubit has a bit flip error ($X$) on it instead of a phase error.  Therefore, if our fault-tolerant circuit has Hadamard gates in it, we need to be prepared to correct both bit flip and phase errors, even if newly-occurring errors are always phase errors.

It is difficult to get past this problem, but there have been some successful efforts to design fault-tolerant protocols for the specific case of noise heavily biased in favor of $Z$ errors (i.e., noise dominated by a dephasing channel).  As in the standard design paradigm, controlling error propagation is paramount, although in this case the goal is to control not just the number but the type of errors, so that phase errors are unlikely to transform into bit flip errors.  This means working with gates which propagate phase errors only into other phase errors.\cite{APdephasing} One gate that has that property is the CNOT gate, but unfortunately, we still must be cautious using CNOT gates in a fault-tolerant protocol for dephasing errors.  This is because, while CNOT propagates existing phase errors to one or two phase errors, if a new phase error occurs \emph{during} the implementation of the CNOT gate, the interaction of the two can result in $X$ or $Y$ errors.  

For CNOTs implemented on qubit Hilbert spaces, this behavior is unavoidable,\cite{noCNOTdephasing} but luckily by going to bosonic codes, there \emph{is} a way around it.  Bosonic codes such as the Kerr cat code\cite{Kerrcat} allow a dephasing-preserving CNOT gate by rotating the state through the extra dimensions of the Hilbert space.  When combined with a code well-suited for correcting phase-biased noise, such as the XZZX code,\cite{XZZX} a variant of the surface code, it is possible to design promising fault-tolerant protocols with improved performance on noise sources dominated by dephasing noise.\cite{XZZXFT}  However, we still do not know how to something similar with more general noise sources.

There are other hardware-specific fault tolerance challenges that will become more and more salient as quantum computers get bigger and start to need fault tolerance.  For example, one common phenomenon in real systems is the presence of cross-talk errors, where performing a gate on one pair of qubits spills over to cause errors on other qubits not involved in the gate.  This is not captured by standard theoretical error models of fault tolerance, and while it is not diffcult to prove that the threshold theorem still holds if there is a reasonable level of cross-talk, it is possible that these errors can noticeably decrease the threshold, making it harder to implement fault tolerance on a machine with such errors.  This may be one of those specific types of error that we will need to design specialized fault-tolerant protocols for, as discussed above.

\section{Fault Tolerance as a Space-Time Code}
\label{sec:spacetime}

The difficulty in making fault-tolerant protocols which prevent specific kinds of errors from changing their nature under propagation suggests it might be wise to look for a new paradigm we can use to design fault-tolerant protocols.  Luckily, there are a number of results in the literature that go beyond the standard paradigm in various ways and together they point to a potential new approach to fault tolerance.

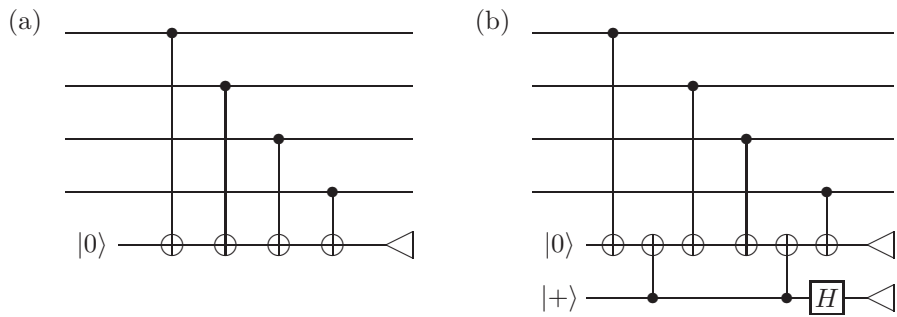
\begin{figure}

\begin{center}
\begin{picture}(330,120)

\put(0,108){\makebox(20,12){(a)}}
\put(45,30){\line(1,0){100}}
\put(25,50){\line(1,0){130}}
\put(25,70){\line(1,0){130}}
\put(25,90){\line(1,0){130}}
\put(25,110){\line(1,0){130}}

\put(25,24){\makebox(20,12){$\ket{0}$}}

\put(65,110){\circle*{4}}
\put(65,110){\line(0,-1){84}}
\put(65,30){\circle{8}}

\put(85,90){\circle*{4}}
\put(85,90){\line(0,-1){64}}
\put(85,30){\circle{8}}

\put(105,70){\circle*{4}}
\put(105,70){\line(0,-1){44}}
\put(105,30){\circle{8}}

\put(125,50){\circle*{4}}
\put(125,50){\line(0,-1){24}}
\put(125,30){\circle{8}}

\put(145,30){\line(2,1){10}}
\put(145,30){\line(2,-1){10}}
\put(155,25){\line(0,1){10}}

\put(175,108){\makebox(20,12){(b)}}
\put(220,10){\line(1,0){84}}
\put(220,30){\line(1,0){105}}
\put(200,50){\line(1,0){135}}
\put(200,70){\line(1,0){135}}
\put(200,90){\line(1,0){135}}
\put(200,110){\line(1,0){135}}

\put(200,24){\makebox(20,12){$\ket{0}$}}
\put(200,4){\makebox(20,12){$\ket{+}$}}

\put(230,110){\circle*{4}}
\put(230,110){\line(0,-1){84}}
\put(230,30){\circle{8}}

\put(245,10){\circle*{4}}
\put(245,10){\line(0,1){24}}
\put(245,30){\circle{8}}

\put(260,90){\circle*{4}}
\put(260,90){\line(0,-1){64}}
\put(260,30){\circle{8}}

\put(280,70){\circle*{4}}
\put(280,70){\line(0,-1){44}}
\put(280,30){\circle{8}}

\put(295,10){\circle*{4}}
\put(295,10){\line(0,1){24}}
\put(295,30){\circle{8}}

\put(310,50){\circle*{4}}
\put(310,50){\line(0,-1){24}}
\put(310,30){\circle{8}}

\put(304,4){\framebox(12,12){$H$}}
\put(316,10){\line(1,0){9}}

\put(325,30){\line(2,1){10}}
\put(325,30){\line(2,-1){10}}
\put(335,25){\line(0,1){10}}

\put(325,10){\line(2,1){10}}
\put(325,10){\line(2,-1){10}}
\put(335,5){\line(0,1){10}}

\end{picture}
\end{center}

\caption{(a) A non-fault-tolerant circuit to measure the eigenvalue of $Z \otimes Z \otimes Z \otimes Z$.  (b) A flag circuit to measure $Z \otimes Z \otimes Z \otimes Z$.}
\label{fig:flags}
\end{figure}

The first thread is flag fault tolerance.\cite{CR17,IBMflags} Consider the circuits in Fig.~\ref{fig:flags}(a) and \ref{fig:flags}(b).  In Fig.~\ref{fig:flags}(a), a phase error on the ancilla qubit after two of the CNOT gates can propagate backwards along the subsequent CNOTs into two phase errors in the qubits of the QECC.  This is a problem, the precise problem we try to avoid by controlling error propagation.  The conventional solution to this is to add extra ancilla qubits and make sure that they interact with different qubits from the code block.  Flag fault tolerance takes a different approach and insteads adds a single extra ancilla as in Fig.~\ref{fig:flags}(b).  In this circuit, a phase error in the same location will still propagate into two qubits of the code block, but will also cause the flag qubit to flip, which is identified when the flag qubit is measured.  The goal is not to control the error propagation, but instead to identify \emph{when} the error occurred.  If we know when the phase error occurred, we will know whether it propagated into a single-qubit error on the code block or a multiple-qubit error.  That is then sufficient to correct it.  The lesson is that it is not necessary to control error propagation if we can identify the precise space-time location where and when the error occurred.

The second thread begins with the idea of code deformation.  One of the standard methods of performing gates for surface codes involves progressively modifying the code to perform some topologically non-trivial transformation on it.\cite{RH,Fowler} There are many other methods of performing fault-tolerant gates that involve switching between codes that are more suitable for one sort of gate or another.\cite{gaugeswitching,JL14,ADP14,BZHJL15} And even transversal gates, apparently so straightforward, can be viewed as a code deformation, dragging the code through a topologically non-trivial loop in the space of all codes.\cite{GZ13}

Standard approaches to code deformation have a fixed QECC as the target, and however we deform the code to perform a logical gate, we always want to return to the original QECC.  This is certainly convenient because it lets us compare the state before the code deformation to the state after the code deformation, but it is also somewhat arbitrary.  Indeed, one can instead repeatedly cycle through a sequence of codes.  None of them is \emph{the} QECC we are using for the protocol; all of them are on equal footing.  This is the idea of a Floquet code.\cite{HH21} The lesson is that the code used in a fault-tolerant protocol can change in time.  Even the Floquet code is too restrictive, as there is no real need to repeat the same sequence of codes every time.

The final thread comes from a result in the model of measurement-based quantum computation (MBQC).\cite{cluster}  In this model, we first prepare a many-qubit entangled state (a \emph{cluster state}) which is independent of the computation to be performed and then perform a sequence of single-qubit measurements, synthesizing the results of the measurements to give the output of the desired quantum computation. Any quantum circuit can be converted into an appropriate sequence of measurements in the MBQC model.  In general, what kind of measurement needs to performed next depends on the outcome to previous measurements, but remarkably this is not true for the sequence of measurements needed for Clifford group circuits.  In particular, all the measurements needed for stabilizer error correction (including fault-tolerant error correction) can be done simultaneously and the outcomes examined to determine the nature of any errors.

The process of converting a quantum circuit into a pattern of measurements in MBQC results in a measurement sequence that is \emph{foliated}, with a sequence of slices each of which corresponds to one moment in time of the original circuit.  This results in a cluster state which has one more dimension than the layout of the original circuit.  For instance, if the circuit involves only nearest-neighbor qubits in two dimensions, the cluster state involves qubits adjacent in three dimensions.  One of the dimensions corresponds to time in the circuit, but if all we are doing is stabilizer error correction, it doesn't matter \emph{which} direction is time, and we can treat the whole thing as a sort of three-dimensional code.\cite{RH} Nickerson and Bombin took this even further by noting that the foliation is arbitrary and unnecessary from the point of view of MBQC.\cite{NB18}  Instead, we can build a fault-tolerant protocol using a cluster state that has no natural foliation.  Even if we want to stay with the circuit model, there is still a lesson we can take to heart here, which is that we should try to treat space and time on as equal a footing as possible.

Putting this all together, what does it give us?  We shouldn't have a fixed QECC, but instead let the code change with time, perhaps not even in a regular cycle.  We should look at our fault-tolerant circuit as a whole, considering space and time together.  Instead of limiting propagation of errors over time and then trying to identify which qubits have errors, we should try to identify the space-time location where an error occurred and perform a correction based on that knowledge.  If we can identify when and where each error occurred and what kind of error it is, it doesn't matter so much how those errors spread or changed afterwards, because we can analyze the circuit ourselves to see how they propagated and what error is in the system now.

Is it even possible to precisely identify where the errors occurred? Perhaps.  Suppose we have a system with $k$ logical qubits and $m$ additional qubits.  Let us look at a section of the circuit, a single gadget, consisting of $T$ gates and suppose we measure $O(m)$ qubits at the end of this gadget.  Presumably we will then reset those qubits, re-entangle them, and continue the protocol from there, but that is not our present concern.  Instead, let us try to determine if it is possible that the $O(m)$ bits we get from the measurement contain enough information to identify the space-time location of every error occurring during the gadget.

Suppose each gate has $a$ possible errors and the error rate per gate is $p$.  Then we expect roughly $pT$ erroneous gates.  There are a total of
\begin{equation}
a^{pT} \binom{T}{pT} \approx 2^{T (h(p) + p \log_2 a)}
\end{equation}
possible sets of faults, where $h(x) = -x \log_2 x - (1-x) \log_2 x$ is the binary entropy function.  There are a total of $k+m$ qubits used in the gadget.  When the gadget has depth $d$ (i.e., consists of $d$ time steps), we have $T = O(d(k+m))$.  We thus need
\begin{equation}
O(d(k+m) (h(p) + p \log_2 a))
\end{equation}
bits of information in order to identify the space-time location where each error originated.  Since $h(p) + p \log_2 a$ is constant independent of the circuit size, when $m = ck$ and the depth $d$ is constant (independent of $k$), then the measurements in principle could have enough information in order to identify the source of every error.

Of course, this analysis is a far cry from actually designing a protocol to do this.  Such a protocol could be analyzed as a spacetime code.  Just as a QECC is structured so that the qubits containing errors can be identified, a fault tolerant protocol would be designed so that the physical and temporal location of errors can be identified.  Some QECCs are \emph{degenerate}, which means that there are some pairs of distinct correctable errors that act the same way on the code space and therefore can't be distinguished, but don't need to be.  In the same way, a traditional fault tolerant protocol would be a degenerate spacetime code, where different temporal locations can't be distinguished because they produce the same overall error.

\section{A Framework for Describing Spacetime Codes}
\label{sec:stspecifics}

\begin{figure}
\begin{center}
\begin{picture}(305,100)

\put(0,88){\makebox(20,12){(a)}}

\put(45,10){\line(1,0){14}}
\put(45,30){\line(1,0){70}}
\put(25,50){\line(1,0){100}}
\put(25,70){\line(1,0){100}}
\put(25,90){\line(1,0){100}}

\put(25,4){\makebox(20,12){$\ket{0}$}}
\put(25,24){\makebox(20,12){$\ket{0}$}}

\put(65,50){\circle*{4}}
\put(65,50){\line(0,-1){24}}
\put(65,30){\circle{8}}

\put(65,90){\circle*{4}}
\put(65,90){\line(0,-1){24}}
\put(65,70){\circle{8}}

\put(59,4){\framebox(12,12){$H$}}
\put(71,10){\line(1,0){54}}

\put(85,10){\circle*{4}}
\put(85,10){\line(0,1){44}}
\put(85,50){\circle{8}}

\put(95,70){\circle*{4}}
\put(95,70){\line(0,-1){44}}
\put(95,30){\circle{8}}

\put(115,30){\line(2,1){10}}
\put(115,30){\line(2,-1){10}}
\put(125,25){\line(0,1){10}}

\put(145,88){\makebox(20,12){(b)}}

\put(190,10){\line(1,0){10}}
\put(190,30){\line(1,0){10}}
\put(170,50){\line(1,0){30}}
\put(170,70){\line(1,0){30}}
\put(170,90){\line(1,0){30}}

\put(170,4){\makebox(20,12){$Z$}}
\put(170,24){\makebox(20,12){$Z$}}

\put(203,10){\circle{6}}
\put(203,30){\circle{6}}
\put(203,50){\circle{6}}
\put(203,70){\circle{6}}
\put(203,90){\circle{6}}

\put(215,85){\line(1,0){10}}
\put(215,75){\line(1,0){10}}
\put(220,85){\circle*{2}}
\put(220,85){\line(0,-1){12}}
\put(220,75){\circle{4}}
\put(213,72){\framebox(14,16){}}

\put(215,45){\line(1,0){10}}
\put(215,35){\line(1,0){10}}
\put(220,45){\circle*{2}}
\put(220,45){\line(0,-1){12}}
\put(220,35){\circle{4}}
\put(213,32){\framebox(14,16){}}

\put(212,10){\line(1,0){4}}
\put(216,6){\framebox(8,8){$\scriptstyle H$}}
\put(224,10){\line(1,0){4}}
\put(210,4){\framebox(20,12){}}

\put(203,73){\line(1,0){10}}
\put(203,87){\line(1,0){10}}
\put(228,73){\line(1,0){9}}
\put(228,87){\line(1,0){9}}

\put(203,33){\line(1,0){10}}
\put(203,47){\line(1,0){10}}
\put(228,33){\line(1,0){9}}
\put(228,47){\line(1,0){9}}

\put(206,10){\line(1,0){4}}
\put(230,10){\line(1,0){3}}

\put(236,10){\circle{6}}
\put(236,30){\circle{6}}
\put(236,50){\circle{6}}
\put(236,70){\circle{6}}
\put(236,90){\circle{6}}

\put(239,90){\line(1,0){27}}

\put(248,25){\line(1,0){10}}
\put(248,15){\line(1,0){10}}
\put(253,15){\circle*{2}}
\put(253,15){\line(0,1){12}}
\put(253,25){\circle{4}}
\put(246,12){\framebox(14,16){}}

\put(248,65){\line(1,0){10}}
\put(248,55){\line(1,0){10}}
\put(253,65){\circle*{2}}
\put(253,65){\line(0,-1){12}}
\put(253,55){\circle{4}}
\put(246,52){\framebox(14,16){}}

\put(236,13){\line(1,0){10}}
\put(261,13){\line(1,0){10}}
\put(238,48){\line(1,-2){10}}
\put(267,48){\line(-1,-2){10}}

\put(236,67){\line(1,0){10}}
\put(261,67){\line(1,0){10}}
\put(238,32){\line(1,2){10}}
\put(267,32){\line(-1,2){10}}

%
%
%

\put(269,10){\circle{6}}
\put(269,30){\circle{6}}
\put(269,50){\circle{6}}
\put(269,70){\circle{6}}
\put(269,90){\circle{6}}

\put(272,10){\line(1,0){20}}
\put(272,30){\line(1,0){10}}
\put(272,50){\line(1,0){20}}
\put(272,70){\line(1,0){20}}
\put(272,90){\line(1,0){20}}

\put(282,24){\makebox(20,12){$Z$}}


\end{picture}
\end{center}
\caption{Constructing a spacetime code from a Clifford circuit. (a) A toy example of a circuit of the form being analyzed.  (b) The corresponding spacetime code, with hollow circles representing qubits for each location (space and time) in the original circuit.}
\label{fig:spacetime}
\end{figure}
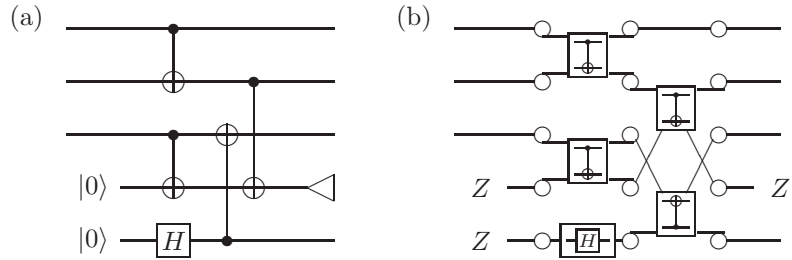

The general structure of a fault-tolerant gadget is illustrated in Fig.~\ref{fig:spacetime}(a).  The gadget takes $n$ qubits as input encoded in some QECC $Q$ and adds $a$ ancilla qubits in the state $\ket{0}$.  We perform a sequence of gates in parallel, but without a hard constraint on the depth of the circuit.  (Higher depths will make it harder to identify and correct all faults, but some parts of the circuit, such as the ancilla preparation, may take a long time and require a large depth.)  Then $n'$ qubits are output from the gadget encoded in a QECC $Q'$ and $b$ qubits are measured in the $Z$ basis.  We don't require $n = n'$ but we do require $n+a = n'+b$, which is the total number of qubits used in the gadget.  For simplicity, we will assume that all ancilla qubits are introduced at the same time and all measurement qubits are measured at the same time.  A more general gadget can be transformed into this form by adding additional waiting steps, which we should consider not to introduce extra errors.  We also allow the number of logical qubits to change from $Q$ to $Q'$, which may involve some new logical qubits being prepared in a standard state or some existing logical qubits being measured and no longer used afterwards.  The one thing that is \emph{not} included in a gadget like this is the ability to condition future operations on measurement results.  Therefore, any fault-tolerant gadget that does that must be broken up into multiple smaller gadgets.  Note that the code $Q'$ may be different from $Q$ and, indeed, which code $Q'$ results from the gadget may depend on the random measurement results.  We allow information about measurement results to be passed classically between gadgets, allowing us to adapt the structure of those later gadgets to earlier events.

If the initial code $Q$ is a stabilizer code and the gadget consists of only Clifford group gates, the output code $Q'$ is a stabilizer code as well.  In this case, we can look at the full stabilizer of the input state, which is composed of elements of the stabilizer $S$ of $Q$ and also the stabilizers $Z$ which constrain the ancilla qubits to be $\ket{0}$.  In a traditional error correction gadget (fault-tolerant or otherwise), the stabilizer elements from $S$ and from the ancilla constraints, when propagated through the circuit, intermingle over the course of the circuit, allowing the propagation of error information into the output measurement qubits.  The stabilizer of the output code $Q'$ (which in this case is the same as $Q$) is formed in the same way from products of elements of $S$ on the input and ancilla constraints which propagate through the gadget to give elements of $S$ on the output.  

The framework also includes cases where we perform a fault-tolerant gate with no error correction.  Some or all measurements may anti-commute with the initial stabilizer ($S$ and ancillas) propagated through the circuit.  The measurements then act to change the code or the logical state.  We can also have output measurements that commute with the initial stabilizer propagated through the circuit but are not themselves members of that set.  A measurement of this form acts as a logical measurement, allowing the construction of fault-tolerant measurement gadgets.  Because we are not requiring $n=n'$, we can also use this framework to describe constructions where the code shrinks or grows.

The next task is to determine a mathematical formalism to describe the fault tolerance properties of the spacetime code.  One strong possibility, which I will discuss here, was given by Bacon {\it et al.,}\cite{spacetime} although I will need to make a couple of modifications to their definition.  Brown and Roberts\cite{BR20} have another possibly relevant framework based on MBQC.  The construction by Bacon {\it et al.}\ takes a Clifford circuit with $\ket{0}$ ancilla states and $Z$-basis measurements and produces a stabilizer subsystem code.  We will have qubits for each physical qubit in the circuit and each time step, as illustrated in Fig.~\ref{fig:spacetime}.  Label the qubits of the spacetime code by two indices $i, t$, with the first index being a qubit number from the original circuit and the second index being the time step.  The time step $t$ goes from $0$ to $T$, which should be even to get the spacetime code to work properly.  If $T$ is odd, we can add an additional time step with no gates in order to make $T$ even.

The gauge group $G$ is defined by elements from four sources.  First, for each single-qubit gate $U$ in the circuit, suppose the gate acts on qubit $i$ between time steps $t$ and $t+1$.   Add two gauge elements, one of the form $X_{i,t} (UXU^\dagger)_{i,t+1}$ and the other of the form $Z_{i,t} (UZU^\dagger)_{i,t+1}$.  These thus encode the error propagation through the gate $U$.  Similarly, for each two-qubit gate $U$ between qubits $i$ and $j$, we add four gauge elements 
\begin{align}
X_{i,t} [U(X \otimes I)U^\dagger]_{(i,t+1),(j,t+1)}, \\
X_{j,t} [U(I \otimes X)U^\dagger]_{(i,t+1),(j,t+1)}, \\
Z_{i,t} [U(Z \otimes I)U^\dagger]_{(i,t+1),(j,t+1)}, \\
Z_{j,t} [U(I \otimes Z)U^\dagger]_{(i,t+1),(j,t+1)}.
\end{align}
Second, for each ancilla $i$ introduced during the circuit at time $0$, we add a gauge element $Z_{i,0}$.  Third, for each $Z$-basis measurement on qubit $i$ at time $T$, we add a gauge element $Z_{i,T}$.  Finally, for each generator $M$ of the stabilizer code which is input to the circuit, add a gauge element $M_{\cdot, 0}$ acting on the appropriate qubits at time $0$.  This last set of gauge generators was not needed by Bacon {\it et al.}\ because they were considering only circuits where the measurements revealed the full error syndrome (thus uniquely specifying the code), but I wish to be more general than that.

The gauge group $G_{st}$ for the spacetime code encodes error propagation through the original circuit as well as constraints on the state in the circuit.  Suppose $P$ is a Pauli group element, perhaps an error, acting on the original circuit at time $t$.  Let $P_t$ be the equivalent Pauli acting on the time $t$ qubits in the spacetime code.  Let $\Pi_{t \rightarrow t'} (P)$ be the result of propagating $P$ through the original circuit from time $t$ to time $t'$ (which could be greater, smaller, or equal to $t$).  Then $P_t$ is equal to $[\Pi_{t \rightarrow t'} (P)]_{t'}$ up to multiplication by elements of the gauge group.  We say $P_t$ and $[\Pi_{t \rightarrow t'} (P)]_{t'}$ are \emph{gauge-equivalent}.

Note that for any $P$, $t \leq T-2$, the product $P_t [\Pi_{t \rightarrow t+1} (P)]_{t+1} [\Pi_{t \rightarrow t+2} (P)]_{t+2}$ is gauge-equivalent to $P_t$, and similarly for any odd product of propagated Paulis.  Following the terminology of Bacon {\it et al.},\cite{spacetime} for $P$ at time $t$, let $\mathrm{spackle}(P) = \prod_{s=0}^T [\Pi_{t \rightarrow s} (P)]_{s}$.  Then, since $T$ is even, $\mathrm{spackle}(P)$ is gauge-equivalent to $P_t$.  

Let $\hat{S}$ be the stabilizer for the input qubits generated by taking the stabilizer $S$ of the input code $Q$ and adding the ancilla constraints $Z$.  It is the full stabilizer of the input state.  Let $\hat{S'}$ be the stabilizer for the output qubits generated by taking the stabilizer $S'$ of the output code $Q'$ and adding the measurement operators $Z$.  Recall that some measurements may act to perform logical operations or to change the code, so it is not necessarily true that $\Pi_{0 \rightarrow T} (\hat{S}) = \hat{S'}$.

If $P$ at time $0$ commutes with $\hat{S}$ and $\Pi_{0 \rightarrow T} (P)$ commutes with $\hat{S'}$, then $\mathrm{spackle}(P)$ commutes with all elements of the gauge group.  This is because $\Pi_{0 \rightarrow t} (P)$ must have the same commutation relation with $Q$ at time $t$ as $\Pi_{0 \rightarrow t+1} (P)$ has with $\Pi_{t \rightarrow t+1} (Q)$.  This means that $\mathrm{spackle}(P)$ has the same commutation relations with the ``input'' and ``output'' Paulis for each gauge generator associated with a gate.  If $P \in \hat{S}$ and $\Pi_{0 \rightarrow T} (P)$ commutes with $\hat{S'}$, or $\Pi_{0 \rightarrow T} (P) \in \hat{S'}$ and $P$ commutes with $\hat{S}$, then $\mathrm{spackle}(P)$ is in the stabilizer $S_{st}$ of the spacetime code.  Otherwise, $\mathrm{spackle}(P)$ that commutes with the whole gauge group is a logical operator.  

Note that if $P \in \hat{S}$ but $\Pi_{0 \rightarrow T} (P) \not\in N(\hat{S'})$ or $P \in \hat{S'}$ but $\Pi_{T \rightarrow 0} (P) \not\in N(\hat{S})$, then $\mathrm{spackle}(P) \not\in S_{st}$ because it does not commute with all gauge generators, even though any $\mathrm{spackle}(P)$ with $P \in \hat{S}$ or $P \in \hat{S'}$ is in $G_{st}$.  In the first case, $P$ is an element of the initial stabilizer (or ancilla constraint) that is replaced by a measurement and no longer applies to the output state.  The second case is when $P$ is the stabilizer element that replaces an initial stabilizer due to a measurement.  This possibility did not exist for Bacon et al.

Next, let us consider how the spacetime code corrects errors and how that reflects the fault tolerance of the circuit.  Let a \emph{fault path} $\kappa$ be a set of locations in the original circuit and Pauli errors associated with those locations.  We will assume that errors occur \emph{after} gates, but before measurements (since after the measurement the qubit is replaced by a classical bit).  That is, a faulty location corresponds to a perfect state preparation or gate followed by an error, or to an error followed by a perfect measurement.  Any error in a gate just before a measurement can be combined with the error associated to the measurement.  The qubits of $Q$ in the incoming state may have errors carrying over into the gadget from previously in the circuit; these errors are assigned to the initial locations on the qubits of $Q$.

The fault path $\kappa$ can then be mapped to an associated Pauli error $K$ on the qubits of the spacetime code and vice-versa.  $K$ is a product of Pauli errors at different times, and these can all be propagated through the circuit to time $t=0$ or alternatively to the final time $T$.  Let $\Pi_{\rightarrow \mathrm{in}} (K)$ be the result of propagating all Pauli errors of $K$ to the initial time $0$ and let $\Pi_{\rightarrow \mathrm{out}} (K)$ be the result of propagating all Pauli errors of $K$ to the final time $T$.  Then $\Pi_{\rightarrow \mathrm{in}} (K)$ and 
$\Pi_{\rightarrow \mathrm{out}} (K)$ are both gauge-equivalent to $K$, and in particular, commute with the same stabilizer elements as $K$.  Let $\Pi_{\rightarrow \mathrm{in}} (\kappa)$ and $\Pi_{\rightarrow \mathrm{out}} (\kappa)$ be the equivalent Pauli errors on the input or output states of the original circuit.  To study whether an error $K$ or the corresponding fault path $\kappa$ is corrected or not, we only need to look at $\Pi_{\rightarrow \mathrm{in}} (K)$ and 
$\Pi_{\rightarrow \mathrm{out}} (K)$, or equivalently at $\Pi_{\rightarrow \mathrm{in}} (\kappa)$ and $\Pi_{\rightarrow \mathrm{out}} (\kappa)$.

And now we need to make one final deviation from Bacon et al.\ by masking some elements of the stabilizer.  The stabilizer elements of $S_{st}$ are derived from initial and final stabilizer elements $M$, and we will define masking based on whether $M$ can be used to give us error information or not.  Let us consider various possibilities:
\begin{arabiclist}
\item $M \in \hat{S}$ and $\Pi_{0 \rightarrow T} (M) \in N(\hat{S'}) \setminus S'$.  These are constraints on the initial state that become logical operators on the final state, representing the case where the circuit is preparing new logical qubits.  We could in principle measure them at this point because the value of the new logical qubit is constrained by the preparation, but the point of preparing new logical qubits is that we want to relax that constraint, meaning these define permanently masked stabilizer elements of $S_{st}$. 

\item $M \in \hat{S'}$ and $\Pi_{T \rightarrow 0} (M) \in N(\hat{S}) \setminus S$.   These are measurements of logical qubits of the initial code.  While they \emph{are} being measured, the measurement tells us information about the encoded state at the start of the circuit and not about the errors.  Therefore, we should also consider $\mathrm{spackle}(M)$ of this form to be permanently masked.

\item $M \in \hat{S}$ and $\Pi_{0 \rightarrow T} (M) \in \hat{S'}$ is a product of $Z$ elements of $\hat{S'}$ corresponding to final measurements.  In this case, we have initial stabilizer elements that are actually being measured by the circuit, so these are always unmasked elements of $S_{st}$.

\item $M \in \hat{S}$ and $\Pi_{0 \rightarrow T} (M) \in \hat{S'}$ which cannot be written as a product of final $Z$ measurements (and so it is a product which includes some non-trivial elements of the stabilizer of $Q'$).  In this case, $M$ is not being measured in this circuit, but because $\Pi_{0 \rightarrow T} (M)$ remains an element of the stabilizer, it may be measured by some future gadget of the fault-tolerant protocol.  In this case, $\mathrm{spackle}(M)$ is temporarily masked.
\end{arabiclist}
Thus, the temporarily masked subgroup is
\begin{equation}
T_{st} = \{ M | M \in \hat{S}, \Pi_{0 \rightarrow T} (M) \in \hat{S'} \}
\end{equation}
and the always unmasked subgroup is
\begin{equation}
U_{st} = \{ M | M \in \hat{S}, \Pi_{0 \rightarrow T} (M) \text{ a product of final $Z$ measurements}\}.
\end{equation}

When we have two possible fault paths $\kappa$ and $\xi$, we can distinguish them via measurements if $\Pi_{\rightarrow \mathrm{out}} (\kappa)$ fails to commute with a different set of measurement $Z$ operators than $\Pi_{\rightarrow \mathrm{out}} (\xi)$, although this is only true for those measurements which actually gain information about the error syndrome rather than those that change the code or logical state.  If $\kappa$ corresponds to spacetime error $K$ and $\xi$ corresponds to spacetime error $L$, then the statement is that $\Pi_{\rightarrow \mathrm{out}} (K)$ and $\Pi_{\rightarrow \mathrm{out}} (L)$ have different syndromes in $U_{st}$.  This is in turn equivalent to saying that $KL \not\in N(U_{st})$.  

There are also cases when we have no need to distinguish $\kappa$ and $\xi$.  One such case is if they are equivalent up to error propagation and multiplication by elements of either $\hat{S}$ or $\hat{S'}$ at the initial and final times, respectively:  Because we only care about the overall error on the ending state, error propagation does not distinguish fault paths.  This means that when $K$ and $L$ are gauge-equivalent via gauge operators associated with the gates, the fault paths are equivalent as well.  If we multiply the final error by a Pauli $M \in \hat{S'}$, then $M$ certainly leaves the state invariant (up to global phase due to errors).  That leaves the case where we multiply the final state by $M$ such that $N = \Pi_{T \rightarrow 0} (M) \in \hat{S}$ but $M \not\in \hat{S'}$.  Since this is gauge-equivalent to multiplying by $N$ at time $0$, it should also leave the state unchanged.  We can understand the behavior at the final time by considering the two cases.  In one case, $M$ is a logical operator for a logical qubit that has just been prepared, and the new logical qubit is constrained to be a $+1$ eigenstate of $M$, so the state remains unchanged.  The other case is when $N$ has been replaced by a new measurement $Z$ on some qubit.  Then $M$ must anticommute with this $Z$ measurement, so $M$ is either $X$ or $Y$ on that qubit.  Therefore applying $M$ changes the measurement result.  However, we need to bear in mind what we do with that measurement result, which is to add $\pm Z$ to the new stabilizer $\hat{S'}$ depending on the measurement outcome.  $M$ changes the measurement result, but it also switches the state into one of the opposite eigenvalue, meaning the state is still correct.  Consequently, we have the following result:

\begin{theorem}
The circuit can correct a set $\mathcal{E}_{FP} = \{\kappa\}$ of fault paths, leaving no residual errors, if and only if the spacetime code can correct the corresponding set of errors $\mathcal{E}_{st} = \{K\}$ using only the always unmasked stabilizer $U_{st}$.  That is, it can correct this set of fault paths or errors iff $KL \not\in N(U_{st}) \setminus G$ for all $K, L \in \mathcal{E}_{st}$.  The circuit can correct all errors from fault paths containing up to $\lfloor (d_U-1)/2 \rfloor$ faults, where $d_U$ is the unmasked distance of the spacetime code.
\end{theorem}

However, this is not the end of the story.  It is not reasonable to expect a gadget to be able to correct all faults that occur during the gadget because there can always be late-occurring faults, such as those on the last layer of gates, that haven't had time to propagate into measured qubits.  Since the gadget is part of a larger fault-tolerant protocol, we can still hope to correct any residual errors in a later gadget.

In order to correct those residual errors later, they need to be distinguishable and not logical errors on the final state.  The residual errors are gauge-equivalent to the spacetime error $K$ corresponding to the fault path $\kappa$, and if they have different error syndromes for the code $Q'$, then there is hope of correcting them later.  If we have $\kappa$ and $\xi$ with residual errors which have the same syndrome for $Q'$ but are not gauge equivalent, then guessing incorrectly as to which is the actual fault path will lead to a logical error.  Therefore, a set of fault paths $\{\kappa\}$ or the corresponding set of errors $\{K\}$ is potentially correctable in the future if $KL \not\in N(T_{st}) \setminus G$.

Unfortunately, this is being overly optimistic.  Whether we can actually correct the residual error depends on what gadgets we do in the future and on the number and nature of any future faults.  To fully understand fault tolerance in this framework, we need to study the protocol as a whole.  Let $\Xi$ be a random variable, a probability distribution on fault paths.  Let $\Xi_{in}$ be a probability distribution on fault paths on the input qubits only and $\Xi_{new}$ be a probability distribution on fault paths restricted to all qubits \emph{except} the input qubits.  Assuming the faults on different gates, state preparation, and measurement locations are independent, then
\begin{equation}
\mathrm{Prob}(\Xi = \xi) = \mathrm{Prob}(\Xi_{in} = \xi_{in}) \mathrm{Prob}(\Xi_{new} = \xi_{new}),
\end{equation} 
where $\xi = \xi_{in} \xi_{new}$ is a fault path composed of $\xi_{in}$ on input qubits and $\xi_{new}$ on all other qubits.  Let $L$ be the spacetime error corresponding to $\xi$ and let $\sigma(L)$ be the unmasked error syndrome of $L$ (i.e., the error syndrome using only generators of $U_{st}$).  We wish to find $\Xi_{out}$ corresponding to $L_{out}$, the distribution of residual errors at the end of our gadget.  This distribution will depend on the measurement results of the circuit.  Any measured qubits that replace stabilizer generators do not give us information about the errors, so in fact, the residual errors depend on $\sigma(L)$.  Since this is information that is available to us and we can in principle adapt our circuit to this information, we will calculate the conditional distribution of $\Xi_{out}$:
\begin{equation}
\mathrm{Prob}(\Xi_{out} = \xi_{out} | \sigma(L) = s) = \mathrm{Prob}(\Pi_{\rightarrow \mathrm{out}} (L) = L_{out})/\mathrm{Prob}(\sigma(L) = s).
\end{equation}
We can then plug in $\mathrm{Prob}(\Xi_{out} = \xi_{out} | \sigma(L) = s)$ as $\mathrm{Prob}(\Xi_{in} = \xi_{in})$ for the next gadget in order to understand the probability of error throughout the full fault-tolerant protocol.

Meanwhile, each gadget has some probability of failing outright, leading to a logical error.  For the actually observed unmasked syndrome $s$, we have a decoding algorithm which deduces some fault path $\kappa_s$ (corresponding to spacetime error $K_s$) which leads to that syndrome and is consistent with syndromes from earlier circuit units.  There is a logical error, as per the analysis above, if for the actual fault path $\xi$ (corresponding to spacetime error $L$), $K_s L \in N(T_{st}) \setminus G$, where $K_s$ and $L$ are the spacetime errors corresponding to $\kappa_s$ and $\xi$.  The probability of this occurring in one gadget, conditioned on $s$, is
\begin{equation}
\mathrm{Prob}(\mathrm{failure}| s) = \left[ \sum_{\xi | \sigma(L) = s \text{ and } K_s L \in N(T_{st}) \setminus G} \mathrm{Prob}(\Xi = \xi) \right] /\mathrm{Prob}(\sigma(L) = s).
\end{equation}
The failure probability accumulates throughout the protocol, so we wish $\mathrm{Prob}(\mathrm{failure}| s)$ to be small for every gadget.  Given any protocol compatible with the spacetime code framework, we can use the above equations to determine if it is actually fault tolerant.

We can understand this analysis in a more qualitative way by deciding on a set of ``acceptable'' residual errors for each gadget, which may depend on the measured syndrome.  All the acceptable errors for a specific measured syndrome of $U_{st}$ must be either gauge-equivalent or have different syndromes for $Q'$ (and thus for $T_{st}$).  The acceptable residual errors become the possible input errors for the next gadget, and we can determine if that gadget is fault-tolerant by looking to see if all likely fault paths of new faults combined with all possible input errors produce one of the acceptable output errors for that gadget.  If so, then the gadget is fault tolerant.  The precise analysis of this will depend on which specific errors are acceptable and how they interact with the likely faults in the circuit, but roughly speaking, the goal is just to make sure that the size of the set of acceptable output errors does not grow from one gadget to the next.  In the standard approach to fault tolerance, this is essentially achieved by insisting that the acceptable residual errors be only errors of low weight.

\section{Conclusion}

The new approach to designing fault-tolerant circuits that I have outlined is more intended as inspiration than a practical approach at this point.  A detailed analysis is certainly possible using the spacetime code, provided we restrict attention to only Pauli noise and Clifford group circuits, but more general noise and circuits may require a more difficult calculation.  More seriously, the need to analyze a full fault-tolerant protocol as a unit is likely impractical.  Instead, we should aim for new heuristics constraining the acceptable residual errors that relax the existing requirements.  Extra freedom to change codes will allow more possible types of fault-tolerant constructions.  While the framework as I have presented it still falls short of a full symmetry between space and time, it does help put them on a more even footing and may point towards new ideas for fault-tolerant gadgets.

\section*{Acknowledgements}

I would like to thank Noah Berthusen, Steve Flammia, Xiaozhen Fu, Jon Nelson, and John Preskill for helpful conversations.


\begin{thebibliography}{99}
\bibitem{NISQ} J.~Preskill, Quantum Computing in the NISQ era and beyond, {\it Quantum} {\bf 2}, 79 (2018); arXiv:1801.00862 [quant-ph].

\bibitem{magicstates} S.~Bravyi and A.~Y.~Kitaev, Universal Quantum Computation with ideal Clifford gates and noisy ancillas, {\it Phys. Rev. A} {\bf 71}, 022316 (2005); arXiv:quant-ph/0403025.

\bibitem{ShorFT} P.~W.~Shor, Fault-tolerant quantum computation, in {\it 37th Symposium on Foundations of Computing (FOCS)}, pp. 56-65 (Burlington, USA, 1996); arXiv:quant-ph/9605011.

\bibitem{chapter} D.~Gottesman, An Introduction to Quantum Error Correction and Fault-Tolerant Quantum Computation, in {\it Quantum Information Science and Its Contributions to Mathematics}, ed.~S.~Lomanaco, Proc. Symp. Applied Math. {\bf 68} (Amer. Math. Soc., 2010), pp. 13-58; arXiv:0904.2557 [quant-ph].

\bibitem{EastinKnill} B.~Eastin and E.~Knill, Restrictions on Transversal Encoded Quantum Gate Sets, {\it Phys. Rev. Lett.} {\bf 102}, 110502 (2009); arXiv:0811.4262.

\bibitem{stabilizer} D.~Gottesman, Class of Quantum Error-Correcting Codes Saturating the Quantum Hamming Bound, {\it Phys. Rev. A} {\bf 54}, 1862 (1996); arXiv:quant-ph/9604038.

\bibitem{CRSS} A.~R.~Calderbank, E.~M.~Rains, P.~W.~Shor, N.~J.~A.~Sloane, Quantum Error Correction and Orthogonal Geometry, {\it Phys. Rev. Lett.} {\bf 78}, 405 (1997); arXiv:quant-ph/9605005.

\bibitem{Poulingauge} D.~Poulin, Stabilizer Formalism for Operator Quantum Error Correction, {\it Phys. Rev. Lett.} {\bf 95}, 230504 (2005); arXiv:quant-ph/0508131.

\bibitem{LDPClocal} N.~Berthusen and D.~Gottesman, work in progress (2022).

\bibitem{GFT} D.~Gottesman, A Theory of Fault-Tolerant Quantum Computation, {\it Phys. Rev. A} {\bf 57}, 127 (1998); arXiv:quant-ph/9702029.

\bibitem{Heisenbergrep} D.~Gottesman, The Heisenberg Representation of Quantum Computers,  in {\it Group22: Proceedings of the XXII International Colloquium on Group Theoretical Methods in Physics}, eds. S.~P.~Corney, R.~Delbourgo, and P.~D.~Jarvis (International Press, 1999), pp. 32-43; longer version arXiv:quant-ph/9807006.

\bibitem{KLZ} E.~Knill, R.~Laflamme, and W.~H.~Zurek, Resilient Quantum Computation, {\it Science} {\bf 279}, 342 (1998).

\bibitem{AB} D.~Aharonov and M.~Ben-Or, Fault-Tolerant Quantum Computation with Constant Error Rate, {\it SIAM J. Comp.} {\bf 38}, 1207 (2008); arXiv:quant-ph/9906129.

\bibitem{KitFT} A.~Y.~Kitaev, Quantum computations: algorithms and error correction, {\it Russian Math. Surveys} {\bf 52}, 1191 (1997).

\bibitem{AGP} P.~Aliferis, D.~Gottesman, J.~Preskill, Quantum accuracy threshold for concatenated distance-3 codes, {\it Quant. Information and Computation} {\bf 6}, 97 (2006); arXiv:quant-ph/0504218.

\bibitem{Knillhighthresh} E.~Knill, Quantum computing with realistically noisy devices, {\it Nature} {\bf 434}, 39 (2005); arXiv:quant-ph/0410199.

\bibitem{DKLP} E.~Dennis, A.~Kitaev, A.~Landahl, and J.~Preskill, Topological quantum memory, {\it J. Math. Phys.} {\bf 43}, 4452 (2002); arXiv:quant-ph/0110143.

\bibitem{RH} R.~Raussendorf and J.~Harrington, Fault-tolerant quantum computation with high threshold in two dimensions, {\it Phys. Rev. Lett.} {\bf 98}, 190504 (2007); arXiv:quant-ph/0610082.

\bibitem{Fowler}  A.~G.~Fowler, A.~M.~Stephens, P.~Groszkowski, High threshold universal quantum computation on the surface code, {\it Phys. Rev. A} {\bf 80}, 052312 (2009); arXiv:0803.0272.

\bibitem{LDPC} D.~Gottesman, Fault-Tolerant Quantum Computation with Constant Overhead, {\it Quant. Information and Computation} {\bf 14}, 1338 (2014); arXiv:1310.2984 [quant-ph].

\bibitem{TZ09} J.-P.~Tillich and G.~Zemor, Quantum LDPC codes with positive rate and minimum distance proportional to $n^{1/2}$, {\it Proc. ISIT 2009}, 799 (Seoul, Korea, 2009); arXiv:0903.0566 [quant-ph].

\bibitem{LTZ15} A.~Leverrier, J.-P.~Tillich and G.~Zemor, Quantume expander codes, {\it 2015 IEEE 56th Annual Symposium on Foundations of Computer Science (FOCS)}, 810 (Berkeley, USA, 2015); arXiv:1504.00822 [quant-ph].

\bibitem{FGL18} O.~Fawzi, A.~Grospellier, and A.~Leverrier, Constant overhead quantum fault-tolerance with quantum expander codes, {\it 2018 IEEE 59th Annual Symposium on Foundations of Computer Science (FOCS)}, 743 (Paris, France, 2018); arXiv:1808.03821 [quant-ph].


\bibitem{GGKL20} A.~Grospellier, L.~Grou{\`e}s, A.~Krishna, and A.~Leverrier, Combining hard and soft decoders for hypergraph product codes, {\it Quantum} {\bf 5}, 432 (2021); arXiv:2004.11199 [quant-ph].


\bibitem{goodLDPC} P.~Panteleev, G.~Kalachev, {\it Asymptotically Good Quantum and Locally Testable Classical LDPC Codes}, in {\it Proc.\ 54th Annual ACM SIGACT Symposium on Theory of Computing (STOC)}, 375 (Rome, Italy, 2022); arXiv:2111.03654 [cs.IT].

\bibitem{Tanner} A.~Leverrier and G.~Z\'{e}mor, Quantum Tanner codes, arXiv:2202.13641 [quant-ph].

\bibitem{GPT22} S.~Gu, C.~A.~Pattison, and E.~Tang, An efficient decoder for a linear distance quantum LDPC code, arXiv:2206.06557 [quant-ph].

\bibitem{DHLV22} I.~Dinur, M.-H.~Hsieh, T.-C.~Lin, and T.~Vidick, Good Quantum LDPC Codes with Linear Time Decoders, arXiv:2206.07750 [quant-ph].


\bibitem{TDB21} M.~A.~Tremblay, N.~Delfosse, and M.~E.~Beverland, Constant-overhead quantum error correction with thin planar connectivity, {\it Phys. Rev. Lett.} {\bf 129}, 050504 (2022); arXiv:2109.14609 [quant-ph].

\bibitem{CKBB21} L.~Z.~Cohen, I.~H.~Kim, S.~D.~Bartlett, and B.~J.~Brown, Low-overhead fault-tolerant quantum computing using long-range connectivity, {\it Sci. Adv.} {\bf 8}, eabn1717 (2022); arXiv:2110.10794 [quant-ph].


\bibitem{KP19} A.~Krishna and D.~Poulin, Fault-tolerant gates on hypergraph product codes, {\it Phys. Rev. X} {\bf 11}, 011023 (2021); arXiv:1909.07424 [quant-ph].

\bibitem{BB22} N.~P.~Breuckmann and S.~Burton, Fold-Transversal Clifford Gates for Quantum Codes, arXiv:2202.06647 [quant-ph].

\bibitem{QWV22} A.~O.~Quintavalle, P.~Webster, and M.~Vasmer, Partitioning qubits in hypergraph product codes to implement logical gates, arXiv:2204.10812 [quant-ph].

\bibitem{timeoverhead} H.~Yamasaki and M.~Koashi, Time-Efficient Constant-Space-Overhead Fault-Tolerant Quantum Computation, arXiv:2207.08826 (2022).


\bibitem{BK21a} N.~Baspin and A.~Krishna, Connectivity constrains quantum codes, {\it Quantum} {\bf 6}, 711 (2022); arXiv:2106.00765 [quant-ph].

\bibitem{BK21b} N.~Baspin and A.~Krishna, Quantifying nonlocality: how outperforming local quantum codes is expensive, {\it Phys. Rev. Lett.} {\bf 129}, 050505 (2022); arXiv:2109.10982 [quant-ph].

\bibitem{DBT21} N.~Delfosse, M.~E.~Beverland, and M.~A.~Tremblay, Bounds on stabilizer measurement circuits and obstructions to local implementations of quantum LDPC codes, arXiv:2109.14599 [quant-ph].

\bibitem{Glocal} D.~Gottesman, Fault-Tolerant Quantum Computation with Local Gates, {\it J. Modern Optics} {\bf 47}, 333 (2000); arXiv:quant-ph/9903099.


\bibitem{GKP} D.~Gottesman, A.~Kitaev, and J.~Preskill, Encoding a Qubit in an Oscillator, {\it Phys. Rev. A} {\bf 64}, 012310 (2001); arXiv:quant-ph/0008040.

\bibitem{cat1} P.~T.~Cochrane, G.~J.~Milburn, and W.~J.~Munro, Macroscopically distinct quantum- superposition states as a bosonic code for amplitude damping, {\it Phys. Rev. A} {\bf 59}, 2631 (1999); arXiv:quant-ph/9809037.

\bibitem{cat2} Z.~Leghtas, G.~Kirchmair, B.~Vlastakis, R.~J.~Schoelkopf, M.~H.~Devoret, and M.~Mirrahimi, Hardware-efficient autonomous quantum memory protection, {\it Phys. Rev. Lett.} {\bf 111}, 120501 (2013); arXiv:1207.0679 [quant-ph].

\bibitem{binomial} M.~H.~Michael, M.~Silveri, R.~T.~Brierley, V.~V.~Albert, J.~Salmilehto, L.~Jiang, and S.~M.~Girvin, New class of quantum error-correcting codes for a bosonic mode, {\it Phys. Rev. X} {\bf 6}, 031006 (2016); arXiv:1602.00008 [quant-ph].

\bibitem{survey} W.-L.~Ma, S.~Puri, R.~J.~Schoelkopf, M.~H.~Devoret, S.~M.~Girvin, and L.~Jiang, Quantum control of bosonic modes with superconducting circuits, {\it Science Bulletin}
{\bf 66}, 1789 (2021); arXiv:2102.09668 [quant-ph].

\bibitem{HK21} Li.~H\"anggli and R.~Koenig, Oscillator-to-oscillator codes do not have a threshold, {\it IEEE Trans. Info. Theory} {\bf 68}, 1068 (2022); arXiv:2102.05545 [quant-ph].

\bibitem{noGaussian} J.~Niset, J.~Fiur\'a{\v s}ek, and N.~J.~Cerf, No-go theorem for Gaussian quantum error correction, {\it Phys. Rev. Lett.} {\bf 102}, 120501 (2009); arXiv:0811.3128 [quant-ph].

\bibitem{OPH+16} N.~Ofek, A.~Petrenko, R.~Heeres, P.~Reinhold, Z.~Leghtas, B.~Vlastakis, Y.~Liu, L.~Frunzio, S.~M.~Girvin, L.~Jiang, M.~Mirrahimi, M.~H.~Devoret, and R.~J.~Schoelkopf, Extending the lifetime of a quantum bit with error correction in superconducting circuits, {\it Nature} {\bf 536}, 441 (2016); arXiv:1602.04768 [quant-ph].

\bibitem {HMC+19}  L.~Hu, Y.~Ma, W.~Cai, X.~Mu, Y.~Xu, W.~Wang, Y.~Wu, H.~Wang, Y.~Song, C.~Zou, S.~M.~Girvin, L.-M.~Duan, and L.~Sun, Demonstration of quantum error correction and universal gate set on a binomial bosonic logical qubit, {\it Nat. Phys.} {\bf 15}, 503 (2019); arXiv:1805.09072 [quant-ph].

\bibitem{CET+20} P.~Campagne-Ibarcq, A.~Eickbusch, S.~Touzard, E.~Zalys-Geller, N.~E.~Frattini, V.~V.~Sivak, P.~Reinhold, S.~Puri, S.~Shankar, R.~J.~Schoelkopf, L.~Frunzio, M.~Mirrahimi, and M.~H.~Devoret, Quantum error correction of a qubit encoded in grid states of an oscillator, {\it Nature} {\bf 584}, 368 (2020); arXiv:1907.12487 [quant-ph].

\bibitem{FNM+19} C.~Fl\"uhmann, T.~L.~Nguyen, M.~Marinelli, V.~Negnevitsky, K.~Mehta, and J.~Home, Encoding a qubit in a trapped-ion mechanical oscillator, {\it Nature} {\bf 566}, 513 (2019); arXiv:1807.01033 [quant-ph].

\bibitem{Terhalboson} C.~Vuillot, H.~Asasi, Y.~Wang, L.~P.~Pryadko, and B.~M.~Terhal, Quantum Error Correction with the Toric-GKP Code, {\it Phys. Rev. A} {\bf 99}, 032344 (2019); arXiv:1810.00047 [quant-ph].

\bibitem{PHboson} J.~Harrington and J.~Preskill, Achievable rates for the Gaussian quantum channel,  {\it Phys. Rev. A} {\bf 64}, 062301 (2001); arXiv:quant-ph/0105058.


\bibitem{APdephasing} P.~Aliferis and J.~Preskill, Fault-tolerant quantum computation against biased noise, {\it Phys. Rev. A} {\bf 78}, 052331 (2008); arXiv:0710.1301 [quant-ph].

\bibitem{noCNOTdephasing} J.~Guillaud and M.~Mirrahimi, Repetition Cat Qubits for Fault-Tolerant Quantum Computation, {\it Phys. Rev. X} {\bf 9}, 041053 (2019); arXiv:1904.09474 [quant-ph].

\bibitem{Kerrcat} S.~Puri, L.~St-Jean, J.~A.~Gross, A.~Grimm, N.~E. Frattini, P.~S.~Iyer, A.~Krishna, S.~Touzard, L.~Jiang, A.~Blais, S.~T.~Flammia, and S.~M.~Girvin, Bias-preserving gates with stabilized cat qubits, {\it Sci. Adv.} {\bf 6}, eaay5901 (2020); arXiv:1905.00450 [quant-ph].

\bibitem{XZZX} J.~P.~Bonilla Ataides, D.~K.~Tuckett, S.~D.~Bartlett, S.~T.~Flammia, and B.~J.~Brown, The XZZX surface code, {\it Nat. Commun.} {\bf 11}, 2172 (2021); arXiv:2009.07851 [quant-ph].

\bibitem{XZZXFT}  A.~S.~Darmawan, B.~J.~Brown, A.~L.~Grimsmo, D.~K.~Tuckett, and S.~Puri, Practical quantum error correction with the XZZX code and Kerr-cat qubits, {\it PRX Quantum} {\bf 2}, 030345 (2021); arXiv:2104.09539 [quant-ph].


\bibitem{CR17} R.~Chao and B.~Reichardt, Fault-tolerant quantum computation with few qubits, {\it npj Quantum Information} {\bf 4}, 42 (2018); arXiv:1705.05365 [quant-ph].

\bibitem{IBMflags} N.~Sundaresan, T.~J.~Yoder, Y.~Kim, M.~Li, E.~H.~Chen, G.~Harper, T.~Thorbeck, A.~W.~Cross, A.~D.~C\'orcoles, and M.~Takita, Matching and maximum likelihood decoding of a multi-round subsystem quantum error correction experiment, arXiv:2203.07205 [quant-ph].


\bibitem{gaugeswitching} A.~Paetznick and B.~W.~Reichardt, Universal fault-tolerant quantum computation with only transversal gates and error correction, {\it Phys. Rev. Lett.} {\bf 111}, 090505 (2013); arXiv:1304.3709 [quant-ph].

\bibitem{JL14} T.~Jochym-O'Connor and R.~Laflamme, Using concatenated quantum codes for universal fault-tolerant quantum gates, {\it Phys. Rev. Lett.} {\bf 112}, 010505 (2014); arXiv:1309.3310 [quant-ph].

\bibitem{ADP14} J.~T.~Anderson, G.~Duclos-Cianci, and D.~Poulin, Fault-tolerant conversion between the Steane and Reed-Muller quantum codes, {\it Phys. Rev. Lett.} {\bf 113}, 080501 (2014); arXiv:1403.2734 [quant-ph].

\bibitem{BZHJL15} T.~A.~Brun, Y.-C.~Zheng, K.-C.~Hsu, J.~Job, C.-Y.~Lai, Teleportation-based Fault-tolerant Quantum Computation in Multi-qubit Large Block Codes, arXiv:1504.03913 [quant-ph].




\bibitem{GZ13} D.~Gottesman and L.~L.~Zhang, Fibre bundle framework for unitary quantum fault tolerance, arXiv:1309.7062 [quant-ph].

\bibitem{HH21} M.~Hastings and J.~Haah, Dynamically Generated Logical Qubits, {\it Quantum} {\bf 5}, 564 (2021); arXiv:2107.02194 [quant-ph].


\bibitem{cluster} R.~Raussendorf and H.~Briegel, A One-Way Quantum Computer, {\it Phys. Rev. Lett.} {\bf 86}, 5188 (2001).

\bibitem{NB18} N.~Nickerson and H.~Bomb\'in, Measurement based fault tolerance beyond foliation, arXiv:1810.09621 [quant-ph].

\bibitem{spacetime} D.~Bacon, S.~T.~Flammia, A.~W.~Harrow, and J.~Shi, Sparse Quantum Codes from Quantum Circuits, {\it IEEE Trans. Info. Theory} {\bf 63}, 2464 (2017); arXiv:1411.3334 [quant-ph].

\bibitem{BR20} B.~J.~Brown and S.~Roberts, Universal fault-tolerant measurement-based quantum computation, {\it Phys. Rev. Research} {\bf 2}, 033305 (2020), arXiv:1811.11780 [quant-ph].

\end{thebibliography}
\end{document}